\documentclass[fleqn,usenatbib]{mnras}

\usepackage{newtxtext,newtxmath,ulem}

\usepackage[T1]{fontenc}
\usepackage{amsmath}
\DeclareRobustCommand{\VAN}[3]{#2}
\let\VANthebibliography\thebibliography
\def\thebibliography{\DeclareRobustCommand{\VAN}[3]{##3}\VANthebibliography}

\usepackage{graphicx}	
\usepackage{amsmath}	

\title[Gamma-rays from the circumgalactic medium of M31]{Gamma-rays from the circumgalactic medium of M31}

\author[Roy et al.]{
Manami Roy,$^{1}$\thanks{E-mail: manamiroy@rri.res.in}
Biman B. Nath,$^{1}$
\\
$^{1}$ Raman Research Institute, 
Sadashiva Nagar, 
Bangalore 560080, India \\ {\rm Accepted for publication in MNRAS on May 19, 2022}}

\date{Accepted XXX. Received YYY; in original form ZZZ}

\pubyear{2022}

\begin{document}
\label{firstpage}
\pagerange{\pageref{firstpage}--\pageref{lastpage}}
\maketitle

\begin{abstract}
We discuss the production of $\gamma$-rays from cosmic rays (CR) in the circumgalactic medium (CGM) of Andromeda (M31) in light of the recent detection of $\gamma$-rays from an annular region of $\sim 5.5-120$ kpc away from the M31 disc. We consider {the CRs accelerated as a result of the} star-formation in the {M31} disk, which are lifted to {the} CGM by advection due to outflow and {CR} diffusion. {T}he advection time scale due to bulk flow of gas triggered by star formation activity in the M31 disc is comparable ($\sim$ Gyr) to the diffusion time scale with diffusion coefficient $\ge10^{29}$ cm$^2$ s$^{-1}$ for the propagation of CR protons with energy {$\sim 412$ GeV} {that are responsible for} the highest energy photons observed. We show that a leptonic origin of the $\gamma$-rays from cosmic ray (CR) electrons {has difficulties,} as the inverse Compton time scale ($\sim$Myr) is much lower than advection time scale ($\sim$Gyr) to reach $120$ kpc. {Invoking} CR electrons accelerated by accretion shocks in the CGM at $\sim100-120$ kpc {does not help since} it would lead to diffuse X-ray features that are not observed.  We, therefore, study the production of $\gamma$-rays via hadronic interaction between CR protons and CGM gas with the help of numerical two-fluid (thermal + CR) hydrodynamical simulation. We find that {a combination of these mechanisms, that are related to the star formation processes in M31 in the last $\sim $ Gyr,} along with diffusion and hadronic interaction, can explain the observed flux from the CGM of M31. 
\end{abstract}




\section{Introduction}
The discovery of $\pi-$meson in 1947 and the consequent $\pi^0\rightarrow2\gamma$ decay pointed towards the fact that $\gamma-$rays can be a useful probe of charged high energetic cosmic ray (CR) particles \citep{Morrison1958}. Unlike CR particles, which gets easily deflected by magnetic field and lose its source information, the $\gamma-$rays retain their directionality. Therefore, sources such as supernova remnants, the Galactic plane and Galactic halo where CR particles are born or confined to, can be observed in $\gamma-$rays. Beginning with the very first $\gamma$-ray telescope carried into orbit, on the Explorer 11 satellite in 1961 to very recent launched Fermi-LAT telescope in 2008, $\gamma$-ray astronomy has come a very long way.  

There have been attempts to observe our nearest neighbour galaxy Andromeda (M31) in $\gamma-$rays since 1970s. Observation by \cite{Ack2017} showed that the inner galaxy (IG) (at galacto-centric distance $<5.5$ kpc) of M31 has $\gamma$-ray luminosity of $(5.0\pm3.0)\times10^{38}$ erg s$^{-1}$ in $0.1-100$ GeV. But a recent analysis of 7.6 years of Fermi-LAT data by \cite{Karwin2019} has revealed a very interesting signature of $\gamma-$ray emission. Their analysis includes observations from a projected distance of $\sim 200$ kpc from M31's centre. They have noticed an excess (positive residual) in the $\gamma-$ray flux after subtracting the contribution from the Milky Way foreground emission as well as  the isotropic component ({\it e.g.} unresolved extra-galactic diffuse $\gamma-$ray emission), residual instrumental back-ground, and possible contributions from other Galactic components with a rough isotropic distribution. In order to explain this excess emission they added M31-related components by considering a uniform spherical template centering the M31, consisting of three regions a) Inner galaxy (IG) : $<5.5$ kpc, b) Spherical halo (SH) : $5.5\hbox{--}120$ kpc c) Outer halo (OH) $120\hbox{--}200$ kpc. They found that the positive residual can be flattened by adding emission from these three spherical regions of M31. They concluded that if this excess originated from M31, it is extended up to $~120-200$ kpc from the centre of M31.  

This observation has triggered a few plausible ideas to explain this extended $\gamma-$ray excess. \cite{Karwin_DM} suggested dark matter annihilation for the origin of this excess. For their model, cold dark matter particles (weakly interacting massive particles (WIMP)) annihilated to bottom quark giving rise to $\gamma-$ray emission. They concluded that DM particles with mass $45\hbox{--}70$ GeV are favourable to produce the observed $\gamma-$ray excess in the SH region. \cite{Rec2021} proposed a different scenario in which the CRs are produced in the galactic centre of M31 and transported to the SH region by means of buoyant bubble(s). They also suggested {\it in-situ} acceleration of CRs at strong shocks in the SH region. These CRs can produce the extended $\gamma-$ray emission through hadronic interaction with protons in the circumgalactic medium (CGM), as well as through inverse Compton scattering of CMB photons by high energy CR electrons. Another recent work by \cite{Zhang2021} considered the  production of CRs due to the star formation activity in the disc and propagation to the SH region by diffusion. The hadronic interaction of those CR protons with CGM protons would give rise to observed $\gamma-$ray emission at SH region. Incidentally, they have not considered advection by outflow for CR transport.

In this paper, we study the case of CRs being produced as a consequence of star-formation activity in the disc and {\it in-situ} acceleration in the shocks of outflowing gas. {Our approach is different from the previous studies in that we consider the CR particles {to be} lifted to the SH region by the combination of advection {via} the outflow {as well as} diffusion. Then they interact with CGM protons and give rise to {the} observed $\gamma-$ray signature in the SH region.} We use two fluids (thermal component and non-thermal component: cosmic ray) hydrodynamical numerical simulation {\it PLUTO} \citep{PLUTO2007}, in order to simulate a $\gamma$-ray image of M31 as observed from the {solar position of} Milky Way and compare with the above mentioned $\gamma-$ray observation. 

\section{Leptonic Interaction}
{Consider first the case of leptonic origin for the observed $\gamma-$ray excess, where CMB photon can give rise to $\gamma-$rays by inverse Compton (IC) scattering by CR electrons. However the IC loss time scale is short (a few Myr), making it difficult to use CR electrons produced in the  galactic centre or disk for the purpose of $\gamma$-ray production at a distance of $\sim 120$ kpc.  \cite{Recchia2021} explored a scenario in which accretion shock due to infalling material can produce the necessary CRs by {\it in-situ} acceleration. This necessiates a shock speed of $\approx 400$ km s$^{-1}$, and gas density, $\approx 10^{-4}$ cm$^{-3}$. However, this scenario poses a problem. Such a shock would also produce diffuse soft X-ray emission, since the post-shock temperature is $\approx 3 \times 10^6$ K. Considering the shocked gas region (width) to be {$\Delta r\sim r_{\rm kpc}$ kpc}, a shell of {radius $r\sim 100$ kpc has volume $4 \pi r^2 \Delta r\approx 3.4 \times 10^{69} \, (\Delta r_{\rm kpc})$} cm$^3$ and would subtend a solid {angle of $2 \pi r\Delta r/D^2\approx 10^{-3} (\Delta r_{\rm kpc}) $} sr. The free-free surface brightness, in $\sim 0.2\hbox{--}1$ keV band, is $\approx 10^{-6} $ erg s$^{-1}$ cm$^{-2}$  sr$^{-1}$. {The post-shock density is assumed to be $4$ times the ambient, since the cooling time is $\approx 3$ Gyr, and the shock is, therefore, not radiative. Note that the surface brightness estimate is independent of the width $\Delta r$ because it appears in the expressions for  both volume and solid angle.}The corresponding count rate in the ROSAT PSPC would be\footnote{using https://heasarc.gsfc.nasa.gov/cgi-bin/Tools/w3pimms/w3pimms.pl} $2.5 \times 10^{-3}$ counts s$^{-1}$ arcmin$^{-2}$, and therefore would have been rather bright in ROSAT all-sky survey in the region, but there is no such feature in the survey map \citep{Snowden1997} (their figure 2f). Therefore, leptonic origin of the $\gamma-$ ray excess does not seem to be a plausible scenario. This prods us to focus on the hadronic origin for the observed $\gamma$-ray. 
}
\section{Hadronic interaction}
{In the hadronic interaction, a fraction $\sim 0.17$ of proton energy goes to pion (e.g, \cite{Rey2008}), which then decays into two $\gamma-$ray photons. For the production of $35$ GeV $\gamma-$rays, {the highest energy photon observed in the $\gamma-$ray excess}, the proton energy should be at least {$35\times2/0.17\,\rm GeV=412\, \rm GeV$}. The source of such CR protons, responsible for the observed $\gamma$-ray excess, can be either the star formation (SF) activity in the disc of M31, or {\it in-situ} formation of CRs in the CGM due to shocks present there. We consider these two possibilities below.
}

\subsection{Star formation activity}
{Considering the mechanical power associated with supernovae (assuming Salpeter initial mass function and stellar masses in the range of $0.1\hbox{--}100$ M$_\odot$) }
and assuming a fraction $\eta\approx 0.1$ of this power being deposited in CRs,  the CR luminosity is given by,
\begin{equation}
\rm L_{\rm CR} 
=1.12\times 10^{41} \, {\rm erg} \, {\rm s}^{-1}\, \frac{\eta}{0.1}  \times \frac{SFR}{(4.8\,\rm M_\odot/\rm year)} \,.
\end{equation} 
As the sound crossing time is $\approx 1$ Gyr for a $10^6$ K gaseous halo of 120 kpc radius, any activities in the disk before 1 Gyr would have left the region of 120 kpc radius by now. Therefore, we consider that the star formation history of the last $1$ Gyr of M31, as given in \cite{Williams2017} (tabulated 1st, 2nd and 3rd columns in table 3 : 3rd column should be multiplied by 3 for scaling to total M31 star-formation rate) would have an effect on the observed $\gamma-$ray excess at SH. Here, we have scaled the star formation rate (SFR) in terms of $4.8$ $\rm M_\odot$ /year, which is the highest value in the last 1 Gyr.  
The CR protons are accelerated in the central region and advected to the CGM by galactic outflow. The advection time scale to a distance 120 kpc (the outer boundary of SH) is 
\begin{equation} 
\begin{split}
t_{\rm adv}&= {\Bigl (\frac{154\pi}{125} \Bigr )}^{1/3} \,  {\Bigl (\frac{R^5\rho_0}{\rm L} \Bigr )}^{1/3} \\
&\approx 1.0  \, \text{Gyr}  {\Bigl (\frac{R}{120 \, \text{kpc}}\Bigr )}^{5/3}  \,
{\Bigl (\frac{SFR}{4.8\, \rm M_\odot/yr}  
\Bigr )} ^{-1/3} 
\, n_{-3}^{1/3}.
\end{split}
\label{eq:adv}
\end{equation} 
Here, the ambient density is written as $n=10^{-3} \, n_{-3}$ cm$^{-3}$. However, the diffusion time scale for protons responsible for 20 GeV photons to reach a distance of $\sim 120$ kpc is ($\approx R^2/6D$)

\begin{equation}
t_{\rm diff}\approx1.0\,\rm Gyr \, \bigl({\frac{R}{120 \rm kpc}\bigr)}^2 \, \bigl({\frac{D_{1\,\rm GeV}}{10^{29} \rm cm^2\,s^{-1}}\bigr)}^{-1} \, \bigl({\frac{E}{412\,\rm GeV}\bigr)}^{-1/3}\,.
\end{equation}
Therefore, if we use the star formation history of M31 in the last $\sim 1$ Gyr, then the 
CRs accelerated in this period would be confined within the SH region. 
Also, the above estimate shows that diffusion can compete with advection if the GeV scale diffusion coefficient is $\ge 10^{29}$ cm$^2$ s$^{-1}$. {However, note that if the diffusion coefficient is smaller than this, say $10^{28}$ cm$^2$ s$^{-1}$, the diffusion time scale increases to $10$ Gyr and can {not compete} with advection, as we will also {show} in our simulation below.}

However, the time scale for pion production by interaction between CR proton and CGM proton is t$_\pi= \frac{1}{\sigma_{pp}\,n_H\,c}=30 \times \bigl({\frac{n}{10^{-3}}\bigr)}^{-1}$ Gyr, which is larger than the escape (diffusion) time. Therefore, M31 (and its CGM) cannot be  a considered a calorimeter for CR protons. The calorimetric fraction (f$_{\rm cal}$(E)) would be $1-e^{-(t_{res}/t_{\pi})}\sim0.04$ (considering diffusion coefficient at 1 GeV to be $10^{29}\rm cm^2\,s^{-1}$). Incidentally, \cite{Krum2020} has recently developed a model for the calometric fraction, by balancing the CR streaming instability and ion-neutral damping. For Milky Way (with similar order of magnitude of star formation rate as in M31), they have estimated f$_{\rm cal}$(E) to be $\sim0.04-0.05$ in the case of 100 GeV CR protons, similar to our estimate derived above. 

Roughly a fraction of $\approx 0.3$ of CR energy is emitted in $\gamma$-rays through $\pi_0$ decay. Hence the expected $\gamma-$ray luminosity is  
\begin{equation}
\label{hadro}
\rm L_{\gamma,H} 
=1.86 \times 10^{39} \, {\rm erg} \, {\rm s}^{-1} \, \Bigl (\frac{\eta}{0.1} \Bigr ) \Bigl (\frac{f_{\rm cal}}{0.05} \Bigr )\frac{SFR}{(4.8 \, \rm M_\odot/\rm yr)}
\end{equation} 
This estimate matches with the calculated the total $\gamma-$ ray luminosity from SH by \cite{Recchia2021} which is $1.7\times 10^{39}$ and  $1.9\times 10^{39}$ erg/sec by using the spectral fit of power-law with exponential cut-off and power law respectively obtained by \cite{Karwin2019}. Note that we have taken the highest SFR over the past 1 Gyr for this estimate.
Consideration of the true SF history can change these calculations a little bit. These estimates are to motivate the idea that 
it is indeed possible to create CR protons by the SF activity that can give rise to observed $\gamma-$ray excess in the SH region.   


 \subsection{In-situ formation}
Next, we explore the case of CR protons  acceleration at a spherical shock located at 120 kpc i.e. the position of SH of M31, ultimately giving rise to 35 GeV $\gamma-$ray photons. For this to occur, the acceleration time scale should be smaller than the pion decay time scale. The acceleration time scale is $ar_L c/(3\times v_{sh}^2$)) \citep{Drury1983} (assuming Bohm diffusion), where $r_L$ refers to the Larmor radius and $a\sim 10$ is a numerical factor that takes into account the compression of gas and magnetic field behind the shock, $v_{\rm sh}$ is shock speed, $B=10^{-6} B_{\mu G}$ is the magnetic field.
Equating this with the  pion-decay loss time for protons, one gets a maximum energy of protons as,
\begin{equation}
E_{max}\sim 2.4 \times 10^8 \, {\rm GeV} \, B_{\mu {\rm G}}\, \Bigl ( \frac{v_{\rm sh}}{100\,{\rm km} \, {\rm s}^{-1}} \Bigr )^2 \, n_{-3}^{-1}\, \Bigl ( \frac{a}{10} \Bigr )^{-1}\,.
\end{equation}
The maximum CR proton energy can  also be calculated according to the Hillas criterion \citep{Hillas1984}, according to which the accelerating region ($L$) should be larger than $2r_L(c/v_{\rm sh})$. 
This yields,
\begin{equation} 
E_{\rm max}\approx \frac{1}{2} \, 10^6 \,{\rm GeV} \, L_{\rm kpc}\, B_{\mu {\rm G}} \, \Bigl ( \frac{v_{\rm sh}}{100\,{\rm km} \, {\rm s}^{-1}} \Bigr )\,,
\end{equation}
where $L_{\rm kpc}$ is the size of the  acceleration region in kpc. These estimates imply that protons with energy $412$ GeV (corresponding to 35 GeV $\gamma$-ray photons) can be produced at 120 kpc i.e. in the SH region. The equipartition B field is of order $0.01 \, \,\mu$G, if the ambient thermal pressure is $\sim 10^{-14}$ erg cm$^{-3}$ (corresponding to, say, $T\sim 10^6$K and $n\sim 10^{-4}$ cm$^{-3}$). Therefore, {\it in-situ} acceleration of CR protons to $\sim 412$ GeV that is required to explain the observations is possible even if the size of the accelerating region is $\sim 80$ pc.

Suppose a fraction $f$ of the shock energy density ($\rho$v$_{\rm sh}^2$, where the velocity of shock at that instant is $v_{\rm sh}= 3R/5t$.), is transferred to CR protons. The CR energy density is then given by
\begin{equation}\label{eq:sw_ecr}
\begin{split}
\epsilon_{CR} &\approx7.7 \times 10^{-15} \text{erg  cm$^{-3}$}  {\Bigl (\frac{R}{120\, \text{kpc}}\Bigr )}^{-4/3}  {\Bigl (\frac{SFR}{4.8\,M_\odot/yr} \Bigr )} ^{2/3} \\ 
&\times n_{-3}^{1/3}\times \Bigl (\frac{f}{0.1}\Bigr ).
\end{split}
\end{equation}

The $\gamma-$ray luminosity due to hadronic interaction between CR proton and CGM proton is
\begin{equation}\label{eq:q_gamma}
\begin{split}
L_{\rm \gamma} &= \int E_{\rm \gamma}\, {\tilde{q}_{\rm \gamma}(E_{\rm \gamma}) }  dE_{\rm \gamma} \int  dV\,  \Bigl [ n_{\rm CGM}  \,  \epsilon_{\rm cr}  \,  \, \, 
\Bigr ] \,,
\end{split}
\end{equation}
where the source function $\tilde{q}$ is given by,
\begin{equation}
\tilde{q}_{\rm \gamma}=\left[\frac{ \sigma_{\rm pp} c\,\left(\frac{E_{\pi^{0}}}{\rm GeV}\right)^{-\zeta_{\rm \gamma}} \left[\left(\frac{2E_{\rm \gamma}}{E_{\rm \pi^{0}}}\right)^{\delta_{\rm \gamma}}+\left(\frac{2E_{\rm \gamma}}{E_{\rm \pi^{0}}}\right)^{-\delta_{\rm \gamma}}\right]^{-\zeta_{\gamma}/\delta_{\gamma}}}{ \xi^{\zeta_{\rm \gamma}-2} \left(\frac{3\zeta_{\rm \gamma}}{4}\right) \frac{E_{\rm p}}{2(\zeta_{\rm p}-1)}\left(\frac{E_{\rm p}}{\rm GeV}\right)^{1-\zeta_{\rm p}}{\it \beta}(\frac{\zeta_{\rm p}-2}{2},\frac{3-\zeta_{\rm p}}{2})}\right]\ .
\label{qt}
\end{equation}
Here $\xi=2$ is the multiplicity factor, $E_{\rm p}/E_{\rm \pi^{0}}$ is the rest mass energy of proton/pions ($\pi^{\rm 0}$), $\zeta_{\rm p}$ and $\zeta_{\rm \gamma}$ are the spectral indices of the incident CR protons and emitted $\gamma$-ray photons respectively, $\delta_{\gamma} = 0.14\zeta_{\rm \gamma}^{-1.6} + 0.44$ is the spectral shape parameter and $ \sigma_{\rm pp}=32(0.96+e^{4.4-2.4\zeta_{\rm \gamma}})$ mbarn (see Equations (8), (19)-(21) in \citealt{Pfrommer2004}). We use $\zeta_p=\zeta_\gamma=2.3$ in our calculations following the spectral fit of \citet{Ackermann2015}.

For CRs produced at the shock front, the corresponding shell volume  is $4\pi R^2 \Delta R 
\approx 5.8\times10^{71}$ cm$^3$, for $R\approx 120$ kpc and $\Delta R\approx 100$ kpc. From  Eq \ref{eq:q_gamma},  the $\gamma$ ray luminosity within the energy range of $0.1\hbox{--}100$ GeV is given by, 
\begin{equation}\label{eq:gamma}
\begin{split}
L_\gamma &=4.5 \times 10^{38} \text{erg/sec}  {\Bigl (\frac{R}{120 \text{kpc}}\Bigr )}^{-4/3}  {\Bigl (\frac{SFR}{4.8\, \rm M_\odot/yr}  \Bigr )} ^{2/3} \\ 
&\times n_{-3}^{1/3}\times \Bigl (\frac{f}{0.1}\Bigr ).
\end{split}
\end{equation}
This is $\sim25$\% of the observed $\gamma-$ray luminosity.   
However, the shock speed  $120$ kpc is 
\begin{equation}\label{eq:sw_velocity}
\begin{split}
v  \sim 67 \, \text{km/s}  {\Bigl (\frac{R}{120 \text{kpc}}\Bigr )}^{-2/3}  \,
{\Bigl (\frac{SFR}{4.8\, \rm M_\odot/yr}  \Bigr )} ^{1/3} n_{-3}^{-1/3}.
\end{split}
\end{equation}
which is roughly equal to the local sound speed for CGM at virial temperature $\approx 10^6$ K. In other words, the outflow at this distance is subsonic. 
Therefore, although the estimated $\gamma-$ray luminosity is comparable to the observed value, the shock at SH region can not be expected to accelerate CRs. At the same time, there would be shocks due to the formation of cold clouds from cooling of the gas behind the shock front, and their turbulent motion. These shocks can accelerate CR protons, which is, however, difficult to estimate analytically. We therefore perform hydrodynamical simulations to build a complete scenario by capturing this process and estimate the $\gamma$-ray luminosity of M31 CGM.

\begin{figure*}
\includegraphics[width=1.0\textwidth]{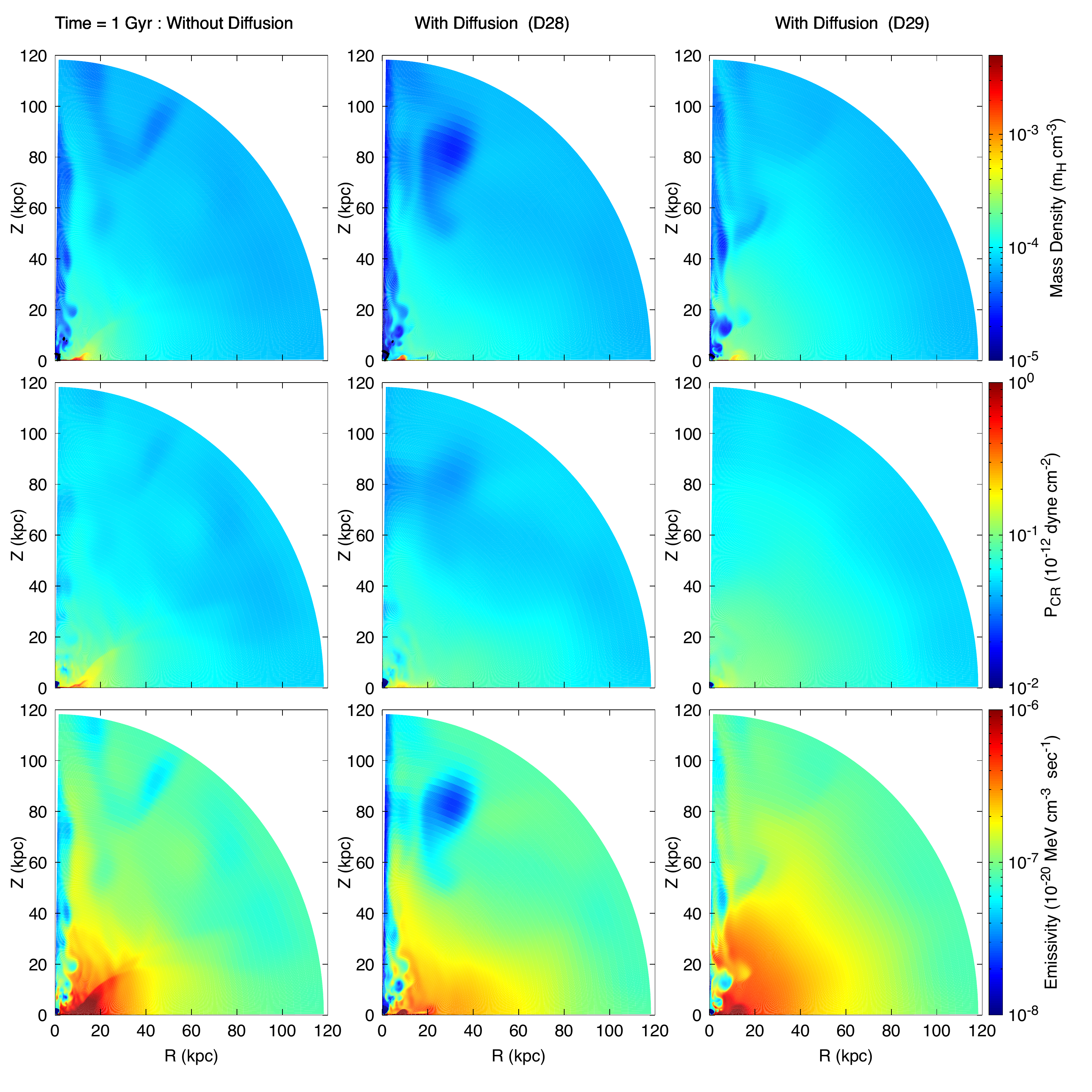}
\caption{{{The density, CR pressure and $\gamma-$ray emissivity at the simulation time of 1 Gyr for the cases of without diffusion (left) and with diffusion : D28 (middle) and D29 (right). Black dots show the shock locations where we inject {\it in-situ} CR particles. }}}
\label{profile}
\end{figure*}

\begin{figure*}
\includegraphics[width=0.45\textwidth]{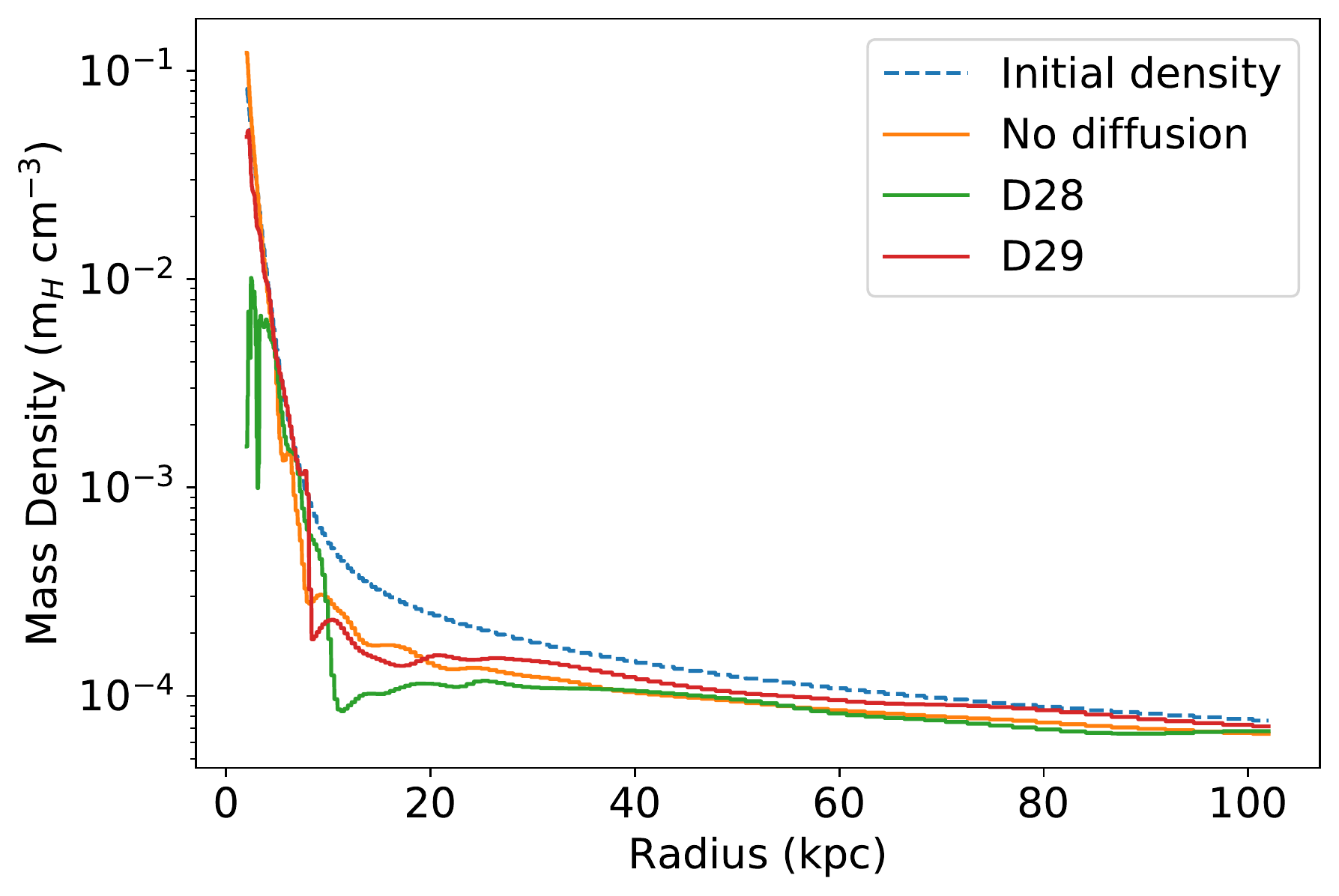}
\includegraphics[width=0.45\textwidth]{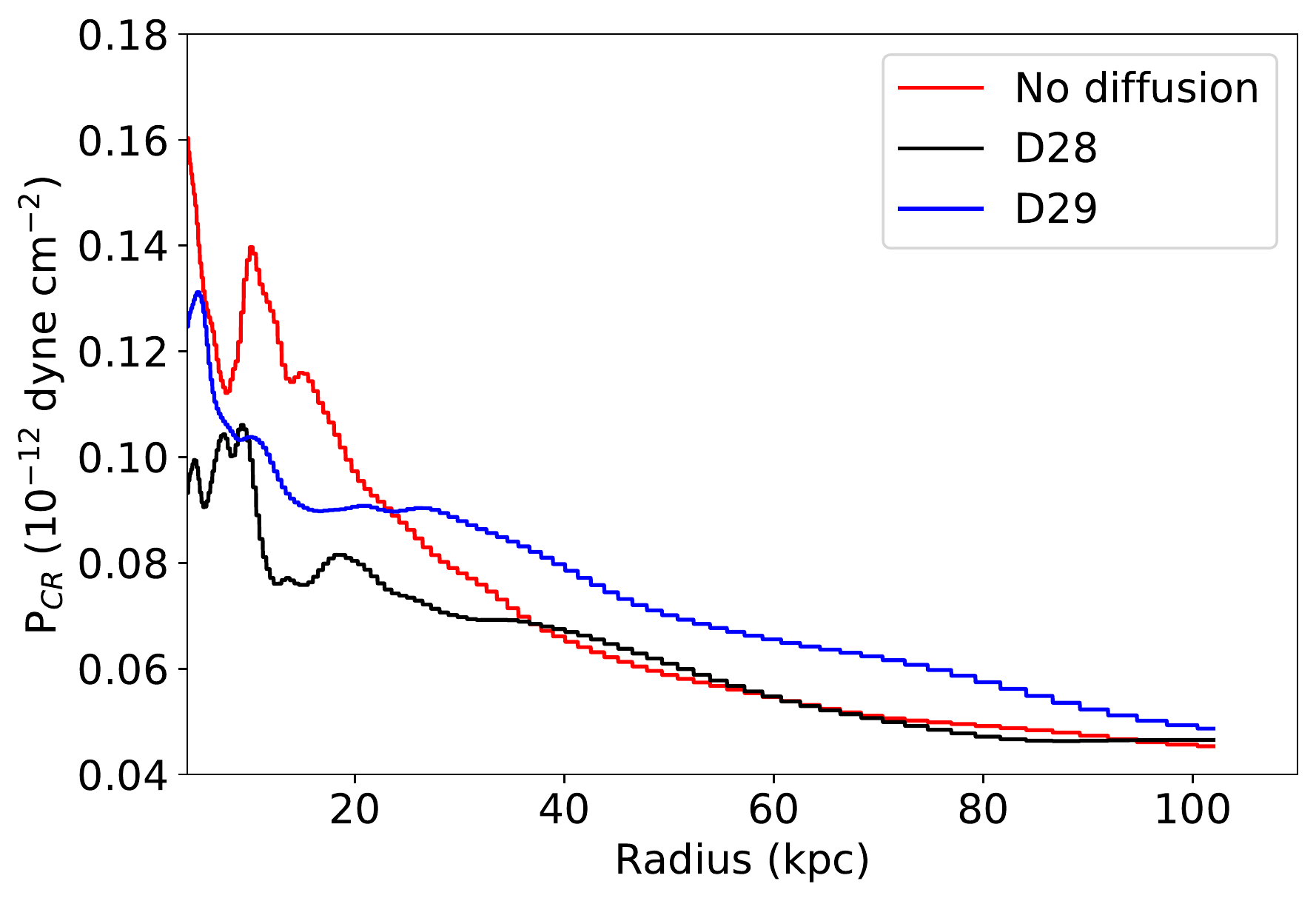}
\includegraphics[width=0.45\textwidth]{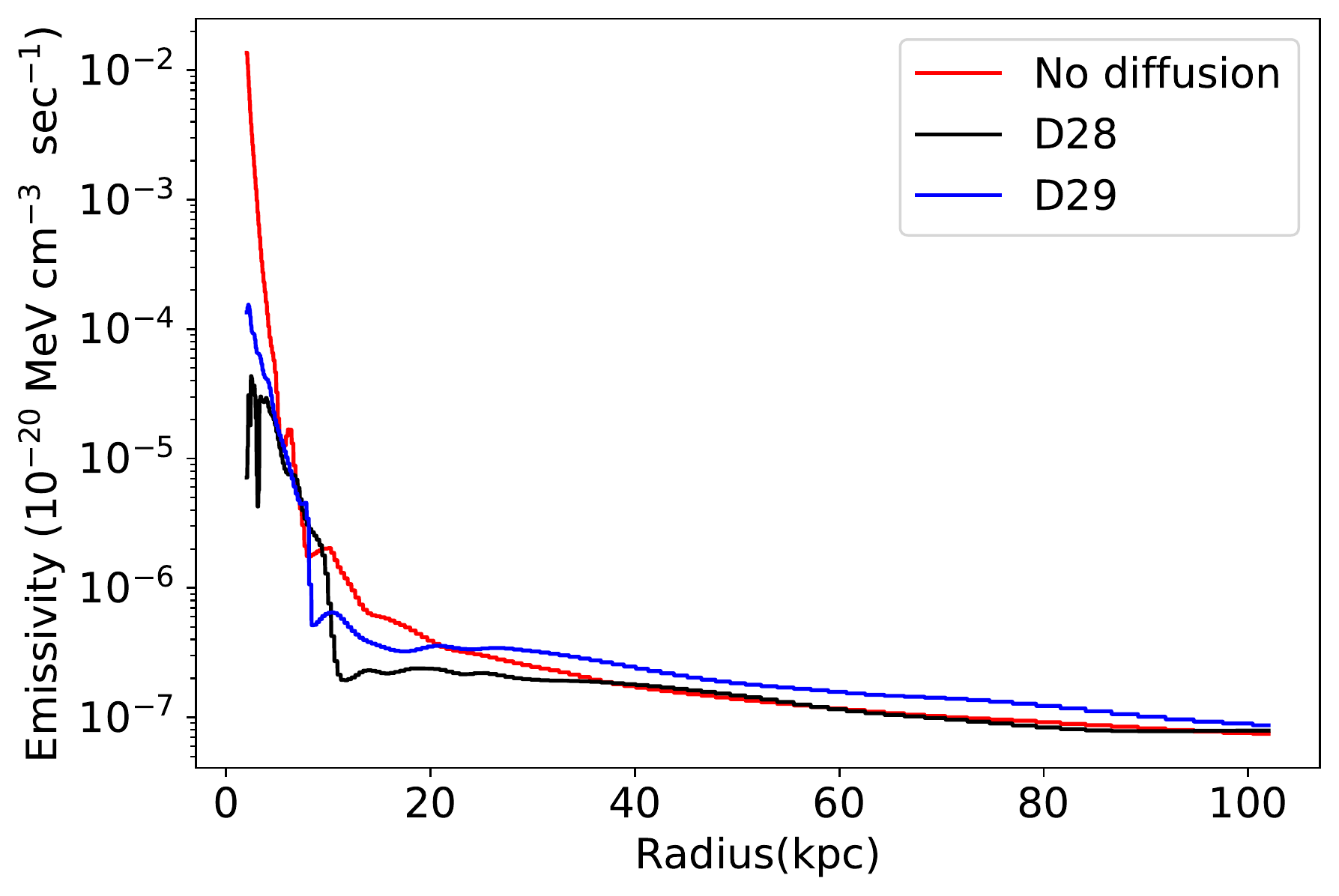}
\includegraphics[width=0.45\textwidth]{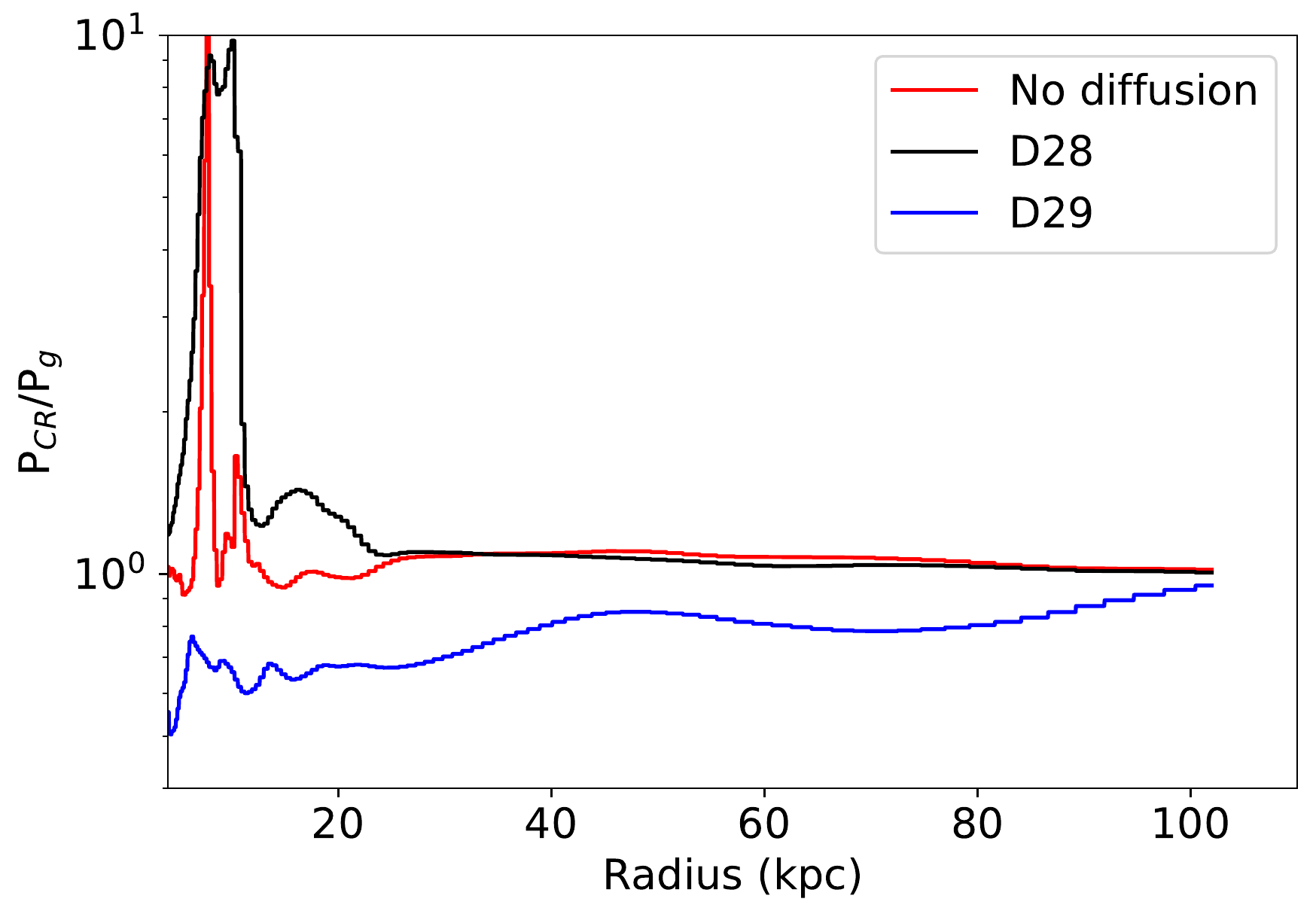}
\caption{{{The radial profiles of volume averaged values of density (Top left panel), CR pressure (Top right panel), emissivity (Bottom left panel) and ratio of CR pressure and gas pressure (Bottom right panel).} }}
\label{radial}
\end{figure*} 

\section{simulation setup}

\subsection{Initial Condition}
The initial galaxy setup for M31 used here is similar to that of \cite{Kartick_long}. We consider two  components: a halo with gas at  $3\times10^6$ K temperature around a disk of $4\times10^4$K temperature. The halo gas contains 11\% of virial mass whereas the stellar disk consists of 5\% of virial mass \citep{MoWhite}. We take into account a rotating disk gas of solar metallicity and a non-rotating halo gas with 0.1 of solar metallicity under the gravitational potential by stellar disk (Miyamoto \& Nagai potential \cite{Miya}) and dark matter halo (modified Navarro-Frenk-White Model \cite{Navarro1997}). We use the cooling function of \cite{Sutherland} for radiative cooling of the warm disc and hot halo gas, and the metallicity dependence is taken into account by linear interpolation of the cooling function between two metallicities (Z$_\odot$ and 0.1Z$_\odot$). Due to its high density and metallicity, the disk gas will cool faster but the stellar radiation, which is not considered here, will maintain the disk temperature at $10^4$ K. To mimic that effect, we exclude the cooling of disk gas by turning off the cooling in the 15kpc$\times$2kpc box whereas we allow cooling of the injected material.  All the parameters used in the galaxy setup are mentioned in table 1 of \cite{Kartick_long}. In addition, we include cosmic-ray as another fluid component and solve the following two-fluid CR hydrodynamical equations using {\it PLUTO} code \citep{PLUTO2007,Gupta2018} where we initially assume thermal pressure to be in equipartition with CR pressure.   {The equations are as follows:}

{\begin{equation}
\frac{\partial \rho}{\partial t} + {\nabla} . (\rho\,\textbf{\textit{v}}) = S_{\rho}
\end{equation}}
{\begin{equation}
\frac{\partial (\rho\textbf{\textit{v}})}{\partial t} + {\nabla} . (\rho\,\textbf{\textit{v}} \otimes \textbf{\textit{v}}) + \nabla(p_{th}+p_{cr}) + \rho \nabla \Phi_t - \frac{\rho \textit{v}_\phi^2}{R} {{\hat{\textbf{R}}}} = 0
\end{equation}}
{\begin{equation}
\begin{split}
\frac{\partial (\rho \textit{v}^2/2 + e_{th} + e_{cr})}{\partial t} &+ {\nabla} . [( \rho \textit{v}^2/2 + e_{th} + 
e_{cr})\,\textbf{\textit{v}}]  \\
&+  {\nabla} . [(p_{\rm th} + p_{\rm cr})\,\textbf{\textit{v}}] =  S_{\rm th} - q_{\rm cool}
\end{split}
\end{equation}}
{\begin{equation}
    \frac{{\partial e_{\rm cr}}}{\partial t} = -{\nabla.}(e_{\rm cr}\textbf{\textit{v}}) -\nabla. \textbf{F}_{\rm cr,diff} - p_{\rm cr}\,\nabla.\textbf{\textit{v}} + S_{\rm cr}
    \label{energy}
\end{equation}}
 {where $\rho$, $v$, $p_{\rm cr}$ and $p_{\rm th}$  denote density, velocity, CR and thermal pressure respectively. $S_{\rho}$, $S_{th}$ and $S_{cr}$ signify the mass and energy sources. $e_{\rm cr}$ and $e_{\rm th}$ are thermal and CR energy densities respectively. The thermal energy lost by radiative cooling is denoted by $q_{\rm cool}$. In the equation \ref{energy}, first two terms in R.H.S denotes CR enenrgy density change due to advection and diffusion respectively where $F_{\rm cr,diff}$ is the CR flux linked to isotropic diffusion process. The third term in the R.H.S of equation \ref{energy} takes into account the adiabatic losses in the CR energy density.} 

\subsection{Injection Condition}
We consider multiple supernova explosions from a large OB association as a continuous source of mechanical energy and mass in the central region of 60 pc, following the criteria that energy deposition rate should be larger than the cooling rate \citep{Sharma2014}.  The metallicity of the injected material is considered to be the same as disk metallicity. We use the star formation history of the last $1$ Gyr of M31, as given in \cite{Williams2017}, because it is the time in which the sound crosses 120 kpc radius through a medium of $10^6$ K gas and any activity in the center beyond this time would have left its signature out of the simulation box of 120 kpc by this time. Considering  Salpeter initial mas function \citep{Salpeter}, 0.1-100 M$_\odot$ stellar mass range, $10^{51}$ erg/sec energy injection by each supernova and efficiency of heating the gas to be 0.3 \citep{Strick2007}, we calculate the mechanical energy injection  to be 
\begin{equation}
    \dot{E}= 7\times10^{40} \text{erg/sec} \,\, \bigl(\frac{\rm SFR}{\,M_{\odot}/\rm yr}\bigr)
\end{equation}
The mass injection rate is assumed to be $\dot{M}= 0.1$ SFR. \\
We also consider 10\% of supernova energy is channelled to CRs. In addition to this (central injection of CRs), we also consider shock injection of cosmic rays. Shocks are detected following the criteria described in \cite{Gupta2018} and we redistribute the total pressure equally between thermal pressure and cosmic ray pressure whenever a shock is detected. 
\subsection{Solver and resolution}
HLL Riemann solver is used to solve the two-fluid hydrodynamical equations and we consider piece-wise linear spatial reconstruction for all variables. Time evolution has been solved with Runge-Kutta 2nd order scheme.  The simulation is executed in 2D spherical geometry ($r \, \theta$) with three velocity components. We consider the computational box from 40 pc to 120 kpc in $r$ direction with uniform grid up to 110 pc (20 grid points) and logarithmic grid (236 grid points) afterwards. In $\theta$ direction, we consider 0 to $\pi/2$ with uniform 256 grid points. We have also performed higher resolution runs with 512 grid points in both directions for convergence study.

\begin{figure*}
\includegraphics[width=1.0\textwidth]{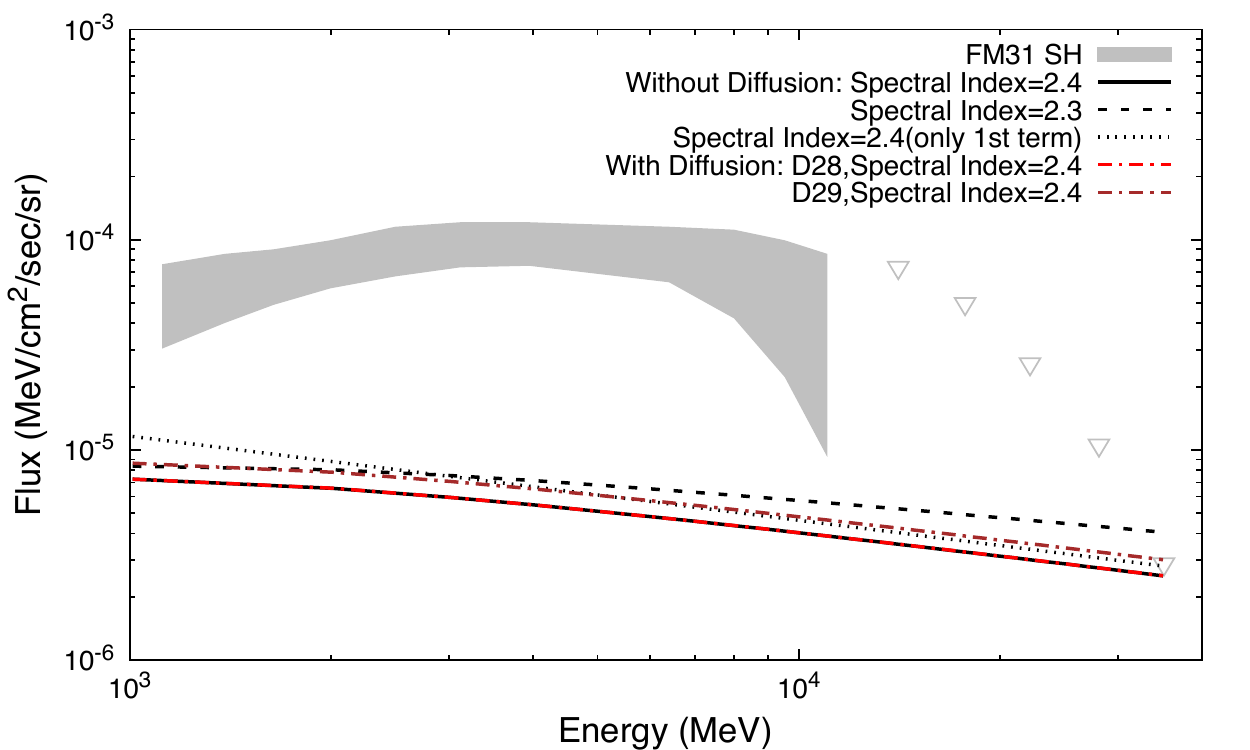}
\caption{The $\gamma-$ray spectra for the cases without and with diffusion  (D28,D29). The variation of $\gamma-$ray flux with differing spectral index is also shown here (black solid, dashed and dotted lines). We also plot the spectra for the case where one consider entirely a negative slope up to 1 GeV. The grey band denotes the observation by \citealt{Karwin2019} including the grey triangles which indicate the upper limits of the observation (Figure 31 of their paper).}
\label{spectrum}
\end{figure*} 
\section{Result}
{In this section, we describe how we use output of our simulation in order to produce $\gamma-$ray map and calculate the $\gamma-$ray flux.} We have used the number densities and CR energy densities from each grid point of the simulation snapshot at 1 Gyr in order to calculate the $\gamma-$ray emissivity of the M31 (eq. \ref{eq:q_gamma} without the volume integral). In Figure \ref{profile}, we have shown the density, CR pressure and $\gamma-$ray emissivity (top to bottom) at snapshot of 1 Gyr with left panel showing no diffusion case, and middle and right panels showing with diffusion cases for $D=10^{28}$ cm$^2$ s$^{-1}$ (D28) and $D=10^{29}$ cm$^2$ s$^{-1}$ (D29). We have indicated the position of shocks in the density plot (top) by black dots where we have injected {\it in-situ} CR particles. Due to low resolution, we do not seem to resolve shocks in the outer CGM and we will discuss this in the last paragraph of this section. The figure shows that the forward shock has faded out at $120$ kpc (as expected from the Mach number estimates in section 3.2). However the advection by outflowing gas has lifted the CRs injected at the central region due to star-formation as well as the {\it in-situ} CRs to $\approx 120$ kpc, the outer region of the simulation box, in 1 Gyr, as anticipated by eqn \ref{eq:adv}. The inclusion of diffusion with D28 does not have much effect with respect to advection as we show above that the diffusion time scale is much larger than the advection time scale in this scenario. However, the enhanced diffusion of CR in the D29 case compared to D28, increases the emissivity of the CGM around M31 as it has comparable time scale as advection time scale, thereby increasing the flux, as we will show below.

 The effect of CR diffusion is clear from the comparison of the leftmost and the rightmost panels, especially in the distribution of CR energy density (middle and bottom columns), which is less patchy and more uniform in the cases with diffusion (D29).  Whereas the resultant emissivity in no-diffusion case is largest surrounding the outflowing gas, along the polar direction, diffusion increases the overall emissivity throughout the CGM.  If we also compare D28 and D29 cases, we can see that in the case of D29, CR energy density is more uniform than the case of D28 : diffusion makes CR distribution smoother in small scale.  An increase in the diffusion coefficient by an order of magnitude increases the diffusive length by $10^{1/2}$, which is reflected in the figure, {\it e.g.} the size of the enhanced CR pressure region above the disk extending up to $\sim 40$ kpc in D28 case, reaches a distance of $\sim 100$ kpc in the D29 case.

\begin{figure*}
\includegraphics[width=1.0\textwidth]{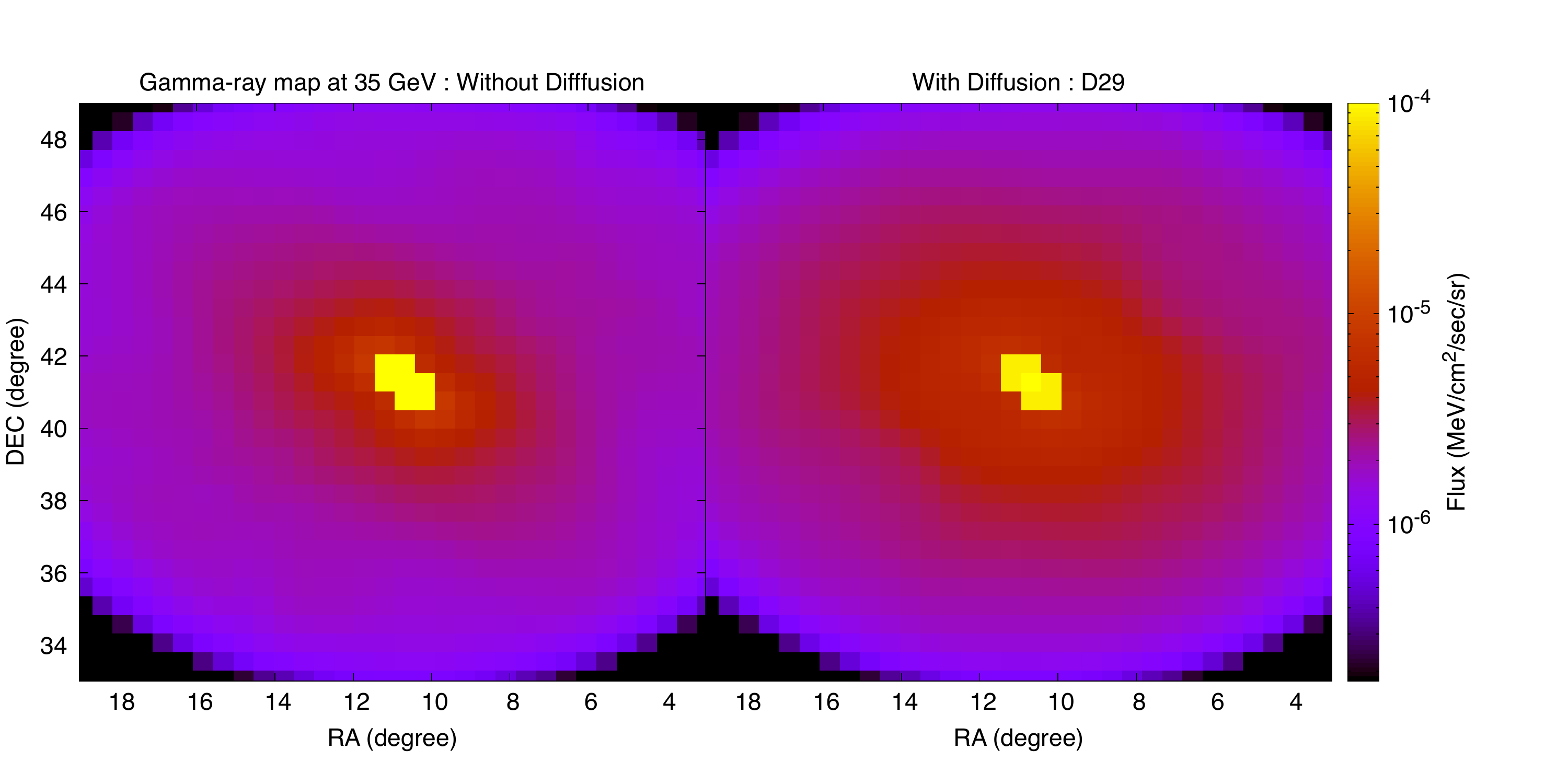}
\caption{{{{Simulated} $\gamma-$ray {image in the sky}  at $35$ GeV  with respect to RA and DEC. We show two cases here : 1)  without any diffusion (left), 2) with diffusion D29 (right). The central point of the map is at the centre of M31 and the yellow portion in the center represents galactic disc of M31 which has a position angle of $37.7^\circ$ and inclination angle of $77.5^\circ$. The {details of the calculation by which the simulation box is rotated, is explained} in the appendix. In the case of without diffusion, the high $\gamma$ flux of  $\approx 10^{-5}$ MeV cm$^{-2}$ s$^{-1}$ sr$^{-1}$ extends up to $2^\circ$, whereas {diffusion (D29)} extends {it} up to $3-4^\circ$.}}}
\label{map}
\end{figure*} 

{We show in Figure \ref{radial} the radial profiles of volume averaged values of density, CR pressure, emissivity and ratio of CR pressure to gas pressure. The top left panel shows the density profile in the initial stage (dashed line), and for the cases of no diffusion and D28 as well as D29. Given the disk geometry, the volume average of annuli at galacto-centric distances $\le 10$ kpc gets diluted in comparison with the disk density. Effectively, the ratio of the average density in the IG to the SH regions turn out to be $\sim 10^{-2}$, with the density in the SH region being $\approx 10^{-4}$ cm$^{-3}$. The CR pressure run would have scaled as $1/r$ in the limit of steady injection of energy (the diffusion coefficient being a constant), but advection makes the profile shallower, by transporting CR into the CGM. These two factors work together to make the ratio of the emissivity in the IG to the SH region $\sim 10^{-3}$. We note that this is consistent with the ratio estimated by \cite{Recchia2021} to explain the surface brightnesses in the IG and SH region. We also note that the emissivity marginally increases with increase in diffusion from D28 to D29. The bottom right panel of Figure \ref{radial} shows the profiles  of the ratio of CR pressure to gas pressure for all the cases which are qualitatively similar to Figure 11 of \citep{Butsky2018}. The ratio in the case of D29 is slightly smaller than D28 and no diffusion case although the CR pressure is larger in D29 than other two cases. This is due to the fact that density hence gas pressure is slightly higher for D29 than D28 and no diffusion case at outer radii. It is reassuring to see that the CR pressure is of the order of gas pressure in the outer halo for all the cases which is also seen in the simulation of \citep{Butsky2018}.  
}

We show the simulated $\gamma-$ray spectrum for different cases in figure \ref{spectrum} for the energy range $1-35$ GeV. We have superimposed the observational data points, corresponding to the power law with exponential cut-off (PLEXP) fit (grey-shaded curve) and upper-limits (grey triangles) from $1$ to $35$ GeV for comparison (Figure 31 of \cite{Karwin2019}). Note that these data points have been obtained after the subtraction of foregrounds from MW and isotropic components, which are model dependent, based on assumed templates. Although \cite{Karwin2019} have demonstrated the robustness of the result, it should be noted that these procedures, each loaded with uncertainty, can introduce systematic errors in the residuals that are larger than shown here. Therefore, instead of trying to explain the exact spectrum of the residuals, we aim to reproduce the flux value within a factor of order unity. Furthermore, since the low energy CRs bring more uncertainty to the calculation as they suffer scattering due to streaming instability which leads adiabatic cooling.  Therefore, we choose $35$ GeV $\gamma-$ray flux to compare with our model, as has also been done by \citep{Zhang2021}. 

Figure \ref{spectrum} shows that the simulated flux matches the observed values at $35$GeV. The flux in the presence of diffusion, with D28,  does not differ from the case of without diffusion, but the flux for the case of D29 significantly increases from the case with no diffusion. The decrease in spectral index from 2.4 to 2.3 also increases flux value. All these parameters are rather uncertain and adjusting them can change the value of the flux, however not significantly. Note that the $\gamma$-ray spectrum has an index similar to the CR spectral index at very high energies, and the spectrum flattens at low energies, near the pion rest mass energy, below which the spectrum has a positive index. We use the full spectrum with these variation, as determined by equation \ref{qt}. Note that \cite{Zhang2021} considered the spectrum to have a negative slope, down to $\sim 1$ GeV, which artificially increases the flux at these energies. 
{We have also a done a similar exercise, and the result is shown with dotted lines in the figure.}


We have considered the M31 coordinates (R.A and DEC) and alignment of M31 such as the position angle, inclination angle  to produce a realistic rotated $\gamma-$ray map (see appendix for the rotation calculations), observed from the position of the solar system. These maps are made with nearly similar resolution of {\it Fermi-Lat} at $35$ GeV $\sim 0.5^{\circ}$. We have considered the $\gamma-$ray spectral index to be $2.4$ and the CR spectral index to be equal to the
$\gamma-$ray spectral index for the production of these maps as well as for the flux calculations. Figure \ref{map} shows the $\gamma-$ray map of M31 at $35$ GeV for the cases of without diffusion (left) and with diffusion: D29 (right). We do not show other map with diffusion D28 as for D28, diffusion does not change the flux in a significant manner as we have seen above. However for D29, we can see there is a significant change from the case of without diffusion. For D29, CRs significantly diffuse out to the outer CGM, making it brighter in $\gamma-$rays. The diffusion D29 increased high intensity region with flux of $\approx 10^{-5}$ MeV cm$^{-2}$ s$^{-1}$ sr$^{-1}$ at 35 GeV from $\approx 2\hbox{--}2.5^\circ$ to $\approx 3\hbox{--}4^\circ$ away from the disc. However, it is difficult to compare the
observed flux with the simulated map by just eye, so for comparison, we have calculated the average flux value with an angular shell of $0.4^\circ\hbox{--}8.5^\circ$ from the centre of M31. The average $\gamma-$ray fluxes calculated for different cases at $35$ GeV are {$2.05\times10^{-6}$, $2.14\times10^{-6}$ and $2.4\times10^{-6}$ MeV cm$^{-2}$ s$^{-1}$ sr$^{-1}$, for no diffusion, D28 and D29 cases respectively}. The fluxes with $D=10^{28}$ cm$^2$ s$^{-1}$ and without diffusion do not differ much, 
whereas, there is 17\% change in $\gamma$-ray flux  with diffusion
coefficient $D=10^{29}$ cm$^2$ s$^{-1}$. {Diffusion increases the emissivity as it allows CRs to interact with a larger volume of CGM, and consequently, a larger number of CGM protons. {To investigate the contribution from the {\it in-situ}, we turned off the injection of energy to CRs in shocks and found a flux $\approx 1.73\times10^{-6}$ MeV cm$^{-2}$ s$^{-1}$ sr$^{-1}$ at 35 GeV. That implies the {\it in-situ} injection contributes $\sim 15\%$ of total flux.}}   

{Doubling the resolution in the case of no-diffusion increases the luminosity from $2.05\times 10^{-6}$ MeV cm$^{-2}$ s$^{-1}$ sr$^{-1}$ to $2.2\times10^{-6}$ MeV cm$^{-2}$ s$^{-1}$ sr$^{-1}$.}
{In other words, a} {two-fold} increase in resolution {increases} the $\gamma-$ray flux by $7-9$\%. {This is due to the fact that the higher resolution in the SH region allows us to resolve more shocks in the CGM, thereby increasing the contribution from the {\it in-situ} CR acceleration.} Therefore, we consider our result as a limit in the sense that the diffusion coefficient considered here can explain the observed results, and larger diffusion coefficient and/or higher resolution may require a smaller CGM mass to explain the same result.


\section{Discussion}
{Previous studies have advocated a variety of proposals to explain the observed $\gamma$-ray excess in the M31 CGM. While \cite{Karwin_DM} invoked the exotic physics of Dark Matter annihilation, \cite{Recchia2021} put forward a scenario of leptonic origin for the $\gamma$-ray and {\it in-situ} acceleration of CR in the accretion shock due to in-falling matter.
The scenario in \cite{Zhang2021}, on the contrary, relies on hadronic $\gamma$-ray production and CR acceleration due to star formation activity in M31 going back to 14 Gyr, but excluding the effect of advection by bulk flow of gas. Our proposal differs from these in that both CR diffusion and advection is shown to be important, with some contribution from {\it in-situ} CR acceleration in shocks in the CGM.}

Our results are roughly consistent with the conclusions drawn by \citep{Salem2016}, who had included CR pressure and diffusion in their cosmological run of galactic winds driving into the CGM of M31-like halos. They found that a diffusion coefficient of $\sim 3 \times 10^{28}$ cm$^2 s^{-1}$ could explain the $\gamma$-ray luminosity of M31. Although theirs was a cosmological simulation from $z=99$ to present day, and with self-regulated star formation in the disc, the concordance is encouraging. In their analytical study, \cite{Zhang2021} had presented a constraint on the baryonic CGM mass of M31 for different values of the diffusion constant, given the $\gamma$-ray luminosity of the SH region. Their limit on the CGM mass (outside 50 kpc) is $\le 5 \times 10^9$ M$_\odot$ for $D_{\rm1 \,GeV}\approx 1.5 \times 10^{29}$ cm$^2$ s$^{-1}$. The limit increases with the increase in the diffusion coefficient, going to $\sim 5 \times 10^{10}$ M$_\odot$ for $D_{\rm 1\, GeV}\sim 1.5 \times 10^{30}$  cm$^{2}$ s$^{-1}$. This is to be compared with the {\it AMIGA} estimate of the M31 CGM mass, of $\ge 4 \times 10^{10}$ M$_{\odot}$ {within the virial radius} \citep{L2020}. {Our setup has CGM mass of $1.4\times10^{10}$ M$_{\odot}$ for the region between $50$ kpc to $120$ kpc.} One marked difference between our approaches is that we have limited our simulations to a look-back time of $1$ Gyr,  {reason for which has been argued above}, whereas their calculation takes into account the star formation history going back to $14$ Gyr, and in particular, an enhanced SFR before a look-back time of $8$ Gyr. The other difference is the lack of advection by bulk flow initiated by star formation process in their calculation. It is possible that these two factors have increased the CGM emissivity for a given CGM mass in their case, and has led to a rather restrictive limit on the CGM mass. {Also note that we have not considered contribution from nuclear activity of the central black hole in M31 which can also enhance the emission. However, that will introduce more free parameters in the calculation which we did not attempt in this paper for the matter of simplicity.}

{\cite{Recchia2021} have argued that the observed $\gamma$-ray intensity at $\sim 100$ Kpc cannot be explained by propagation of CRs produced in the disk, with CR diffusion or advection by galactic outflow. They have first argued that the ratio of emissivities in the IG and SH region is $\sim 10^{-3}$, based on the observed intensities in these regions. Incidentally, this is consistent with the bottom left panel of Figure \ref{radial}. 
However, their discussion of CR galactic outflow is based on assumptions of magnetic field geometry, for stationary wind, and on the assumption of increasing diffusion coefficient with distance. {The profiles of cosmic ray pressure, and its ratio to thermal pressure, from our simulations are qualitatively and quantitatively consistent with other recent simulations that do not make these assumptions, as explained in section 3.}   The distance dependence of diffusion coefficient comes from the {assumption} that if CRs excite waves through streaming instability in the ionized galactic halo, the diffusion coefficient {would be} smaller where the source density is larger  enhancing the CR density in the inner galaxy and it would be larger in the spherical halo region. However this argument {may not be relevant} in our case as streaming instability is dominant in low energy CR protons and does not apply to 400 GeV CR protons which are responsible for 35 GeV $\gamma-ray$ flux that we described in the paper. {In fact,} Figure 10 of \cite{Recchia2016} {clearly shows} that the difference in diffusion {coefficient} with distance decreases as one increases the energy and it is almost negligible at CR energy of 400 GeV. 
These considerations lead us to conclude that CR diffusion and advection can indeed explain the observed $\gamma$-ray emission from the CGM of M31. {If, however, the diffusion coefficient turns out to be increasing with distance as described in \cite{Recchia2021}, then it would be difficult for advection-diffusion processes to explain the observed flux. }
}

In a recent paper, \cite{Blasi2019} have pointed out that if the diffusion coefficient is very large, then the CGM magnetic field of strength $\approx 0.01 \, \mu$ G will be amplified by a CR driven instability, thereby producing a large gradient in CR pressure, and moving the gas by advection with speed $10\hbox{--}100$ km s$^{-1}$. This is one of the reasons we have not explored larger values of diffusion coefficient in our simulation. The close connection between diffusion and advection that is the crux of our result, will be more intimate in this scenario.

Our result of a brighter CGM in the case of enhanced diffusion, with $D$ (1 GeV) $\ge 10^{29}$ cm$^{-2}$ s$^{-1}$, is potentially testable, if the SH region is further divided into two shells and the observed flux from the outer part of the SH shell is compared with simulation result. {Taking the total region between $5.5$ kpc to $120$ kpc as a whole does not allow one to discuss the distribution of flux in this region, whether or not the outer part of SH is as bright as the inner part of SH, which will be needed to test our results.} Our result shows that increasing the diffusion coefficient beyond $10^{29}$ cm${^2}$ s$^{-1}$ would render the intensity of a region $> 3\hbox{--}4^\circ$ away from the disc to be $\approx 10^{-5}$ MeV cm$^{-2}$ s$^{-1}$ sr$^{-1}$ at 35 GeV. Future analysis may be able to test this prediction.

Incidentally, since the scenario advocated here requires CR protons in the CGM of M31, one can ask if the corresponding CR electrons would emit observable synchrotron radiation, assuming the CGM to be magnetized. We can use eqn \ref{eq:q_gamma} to the CR proton energy density corresponding to the observed $\gamma$-ray luminosity. For a volume of $4\pi R^2 \Delta R$ (with $R\approx 120$ kpc and $\Delta R\approx 100$ kpc), the CR proton energy density is 
$1\times10^{-15} \, n_{-3}^{-1}$ erg/cc. Typically, CR electrons carry $\approx 1\%$ of the total CR energy density. This implies 
CR electron energy density of $\epsilon_{cr,e}\approx 10^{-17}\, n_{-3}^{-1}$ erg cm$^{-3}$. For a power law distribution of electron ($N(E)dE=AE^{-p}dE$, with $p=2.5$), the emissivity  at $400$ MHz (eqn 6.36 in \cite{Rybicki1986}) is
$2\times10^{-36}\times B^{1.75} \, n_{-3}^{-1}$ erg s$^{-1}$ cm$^{-3}$ sr$^{-1}$ Hz$^{-1}$. The equipartition value of the magnetic field (with the thermal energy density) is $B\approx 3\mu \, {\rm G} \, n_{-3}^{1/2}\,(T/10^6 \, {\rm K})^{1/2}$. This implies an emissivity at $400$ MHz of $4\times10^{-46} \, n_{-3})^{-0.125}\,(T/10^6 \, {\rm K})^{0.875}$ erg s$^{-1}$ cm$^{-3}$ sr$^{-1}$ Hz$^{-1}$. Therefore, the expected radio flux  is 
$\approx 2.5\, \mu$  Jy per arcmin$^2$, which is difficult to observe. For example, 
the beam size of GMRT telescope for this waveband is $2\times2$ arcmin$^2$, implying a signal in a single beam of 
$8 \mu$ Jy, whereas the noise is roughly $13 \, \mu$ Jy for GMRT. 

\section{Conclusion}

{We investigate the plausible origin of {recently detected} $\gamma-$ray signature in the CGM (at $\sim120$ kpc) of M31 by \cite{Karwin2019} {using} hydrodynamic simulation with two fluid (thermal + CR). We consider CR produced in stellar disk by star-formation activity {that diffuse and also get} advected outwards and produce $\gamma-$ray by hadronic interaction. {We argue that since accoustic wave propagation timescale to $120$ kpc in the CGM gas of} of $10^6$ K gas {is} $\approx 1$ Gyr, any disturbance in the disc before $1$ Gyr would have departed the $120$ kpc halo and has little effect in observed $\gamma-$ray. {We therefore} use {the} star-formation history of M31 over the past $1$ Gyr {in our simulation}.  We also include {\it in-situ} acceleration of CR in shocks in the outflow regions{, as well as CR diffusion}.} {We find that advection and diffusion of CR produced in the M31 disc (due to star formation activity) and in CGM shocks can explain the observed flux, with CR diffusion coefficient (at 1 GeV) $\approx 10^{29}$ cm$^2$ s$^{-1}$, for a CGM mass that is $\approx 10\%$ of the total halo mass of M31. {We estimate the contribution of {\it in-situ} accelerated CR in the CGM  to be of order $\sim 15\%$ towards the $\gamma$-ray luminosity.} Increasing the diffusion coefficient beyond this increases the flux, especially towards the outer parts of the SH region, making a region of $>3\hbox{--4}^\circ$ around the M31 disc shine with a diffuse flux of $\approx 10^{-5}$ MeV s$^{-1}$ cm$^{-2}$ sr$^{-1}$ at $35$ GeV. Our work emphasises the hadronic nature of the observed $\gamma$-ray excess in M31 CGM, as well as the comaparable contribution of advection and diffusion of CR towards CR propagation in the M31 CGM.
}

\section*{Acknowledgements}

We wish to thank Yi Zhang and Chris Karwin for useful discussions. {We also thank the anonymous referee for the interesting comments which have been helpful to improve the manuscript.}

\section*{Data Availability}

The data underlying this article are available in the article.



\bibliographystyle{mnras}
\bibliography{reference} 




\appendix

\section{Geometry}

\begin{figure*}
\includegraphics[width=0.8\textwidth]{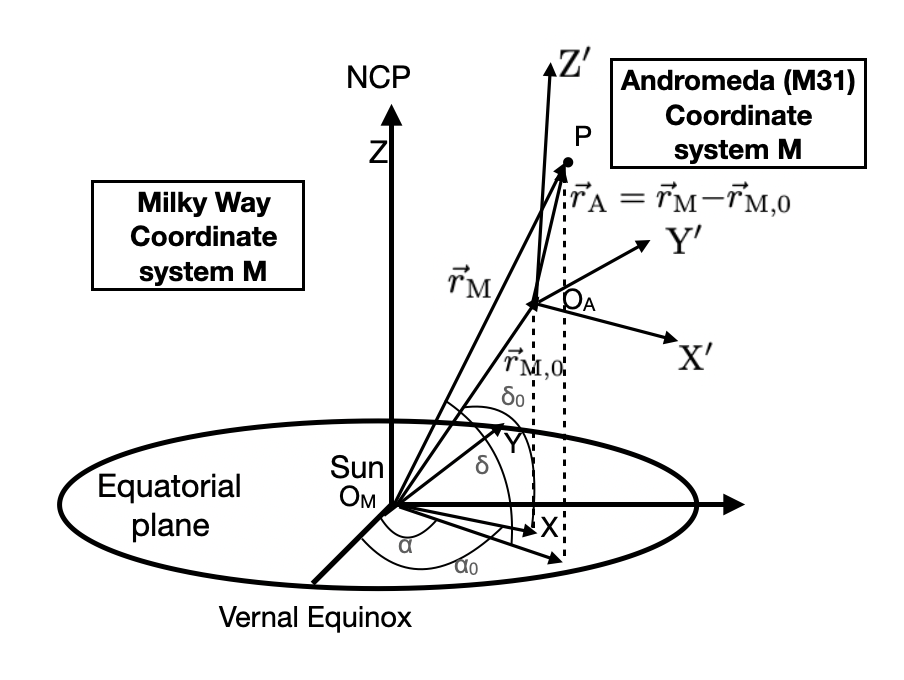}
\caption{{Coordinate systems of M31 and Milky Way}}
\label{geo1}
\end{figure*}

Consider two coordinate systems, one for the Milky Way (MW) and another for Andromeda (M31) (See the Figure \ref{geo1} for reference). We denote the coordinate system for MW by `M' (XYZ), which is centered at the solar position (O$_M$). In this coordinate system, the Z-axis points towards the North celestial pole (NCP), X-axis is directed towards the projection of the centre of M31 in the equatorial plane, and the Y-axis is perpendicular to X-axis in the equatorial plane. The coordinate system for M31 is denoted by `A' (X$^\prime$Y$^\prime$Z$^\prime$) and whose origin is at centre of M31 (O$_A$) with equatorial coordinates, RA$=\alpha_0$ and DEC$=\delta_0$ respectively. Its axes are aligned with the axes of `M'.  Consider any point P in the sky with RA and DEC of  $\alpha,\delta$. Andromeda's orgin O$_A$ and the point P have position vectors $\vec{r}_{M,0}$ and $\vec{r}_M$ respectively w.r.t the coordinate system `M'. Therefore, the position vector of the point P w.r.t  M31 coordinate `A' would be $\vec{r}_A=\vec{r}_M-\vec{r}_{M,0}$ .

The three components of $\vec{r}_M$ is 
\begin{equation}
\begin{split}
&x_M= r_M \cos(\alpha-\alpha_0) \cos(\delta),  \\
&y_M= r_M \sin(\alpha-\alpha_0) \cos(\delta),\,\, \text{and} \\
&z_M= r_M \sin(\delta).
\end{split}
\end{equation}
Similarly, the three components of $\vec{r}_{M,0}$ is
\begin{equation}
\begin{split}
&x_{M,0}= r_{M,0}  \cos(\delta_0),  \\
&y_{M,0}= 0,\,\, \text{and} \\
&z_{M,0}= r_{M,0} \sin(\delta_0).
\end{split}
\end{equation}

However, in reality, `M' (MW) and `A' (M31) coordinates are not aligned . M31 has a position angle (P.A) of $37.5^\circ$ and inclination (i) of $77.5^\circ$. Therefore, to align `M' and `A' coordinates, we need to rotate either of the coordinate systems. In order to take into account P.A and i of M31, we apply three rotations on the position vector $\vec{r}_A$ and obtain the local M31 coordinates corresponding to each line of sight $\vec{r}_M$ w.r.t. the solar position of MW :
\begin{equation}
\begin{split}
(x_A\,\,y_A \,\, z_A)^T= &R_{y,clock}(90-i)\times R_{x,anti}(90-\theta)\times R_{y,clock}(\delta_0) \\
&\times ((x_M-x_{M,0}) \,\, (y_M-y_{M,0})\,\, (z_M-z_{M,0}))^T 
\end{split}
\end{equation}

\begin{equation}
\begin{split}
&r_A = \sqrt{{x_A}^2+{y_A}^2+{z_A}^2} \\
&\theta_A=\arccos(z_A/r_A) \\ 
\end{split}
\end{equation}
\\ \\
where, the rotation matrices are : $R_{x,anti}(\theta)=$                    
$\begin{pmatrix}
1 & 0 & 0\\
0 & \cos \theta  & \sin \theta  \\
0 & -\sin\theta  & \cos\theta  
\end{pmatrix}$ and $R_{y,clock}(\theta)=$                    
$\begin{pmatrix}
\cos \theta & 0 & \sin \theta \\
0 & 1 & 0 \\
-\sin \theta & 0 & \cos\theta 
\end{pmatrix}$

A pair of $\alpha$ and $\delta$ corresponds to a particular line of sight. Each point along such a line of sight from the M31 has a particular line-of-sight distance from the solar position, from which we get a pair $r_A$ and $\theta_A$ in the local M31 coordinate using above method. If $r_A$ lies within our simulation box we find the grid in the simulation box where the coordinate pair ( $r_A$ , $\theta_A$) lies. We then use the corresponding density and CR pressure in that grid to calculate the $\gamma-$ray emissivity. {This is repeated for all points along a given line of sight (corresponding to a given $\alpha, \delta$), and then} integrated it along the line of sight. {The integrated value at each $\alpha, \delta$ then gives} the $\gamma-$ray flux {in that sky coordinate}, in order to produce a $\gamma$-ray image as shown in Figure \ref{map}. 


\bsp	
\label{lastpage}
\end{document}